\documentclass[12pt,preprint]{aastex}
\shorttitle{Element Distributions}
\shortauthors{Satterfield et al.} 
\received{2011 November}
\begin{document}

\title{ELEMENT DISTRIBUTIONS IN THE CRAB NEBULA
\footnote{This paper involves data obtained at the MDM Observatory
and at the McDonald Observatory of The University of Texas at Austin.}
}

\author{Timothy J. Satterfield, Andrea M. Katz, Adam R. Sibley, and Gordon M. MacAlpine}
\affil{Department of Physics and Astronomy, Trinity University, San Antonio, TX 78212}

\and

\author{Alan Uomoto}
\affil{Carnegie Institution for Science, The Observatories, Pasadena, CA 91101}

\begin{abstract}
Images of the Crab Nebula have been obtained through custom interference filters which transmit
emission from the expanding supernova remnant in He~II~$\lambda$4686, H$\beta$, He~I~$\lambda$5876,
$[$O~I$]$~$\lambda\lambda$6300,6364, $[$N~II$]$~$\lambda\lambda$6548,6583, $[$S~II$]$~$\lambda\lambda$6716,6731,
$[$S~III$]$~$\lambda$9069, and $[$C~I$]$~$\lambda\lambda$9823,9850.
We present both raw and flux-calibrated emission-line images.  Arrays of 19,440 photoionization models,
with extensive input abundance ranges, were matched pixel by pixel to the calibrated data
in order to derive corresponding element abundance or mass-fraction distributions for helium, carbon,
nitrogen, oxygen, and sulfur.  These maps show distinctive structure, and they illustrate 
regions of gas in which various stages of nucleosynthesis have apparently occurred, including the CNO-cycle,
helium-burning, carbon-burning, and oxygen-burning.  It is hoped that the calibrated observations 
and chemical abundance distribution maps will be useful for developing a better understanding 
of the precursor star evolution and the supernova explosive process.

\end{abstract}
  

\section{Introduction}

The Crab Nebula (M1 = NGC1952) is the visible remnant
of the core-collapse supernova SN1054. Based on evidence from chemical processing in the line-emitting gas,
MacAlpine \& Satterfield (2008, herein MS) deduced an initial precursor star mass $\gtrsim$~9.5~M$_{\odot}$,
and Wanajo et al. (2009) have suggested that a $\thicksim$10~M$_{\odot}$ precursor underwent an O-Ne-Mg core collapse 
to form the neutron star (PSR~B0531+21) near the center of the nebula.  The visible emitting gas and neutron 
star have a combined mass of at least several M$_{\odot}$ (MacAlpine \& Uomoto 1991), whereas the outer layers 
of the precursor may have been expelled in a presupernova wind (Nomoto et al. 1982) or a shockwave (Chevalier 1977).
Major components of the visible nebula are synchrotron continuum emission produced by the pulsar wind and
thermal line emission of material ejected from the interior of the precursor star.  The expelled thermal gas,
moving outwards with speeds up to about 1700~km~s$^{-1}$ (e.g., MacAlpine et al. 1996), is ionized and heated primarily
by the synchrotron radiation field.  The complex observed filamentary structure is characterized by a range
of nuclear processing (MacAlpine et al. 2007, MS).  Comprehensive reviews of the Crab Nebula have been
prepared by Davidson \& Fesen (1985) and Hester (2008), and a useful discussion of element abundances
was previously given by Henry (1986).

The Crab Nebula is particularly well suited for investigating nucleosynthesis processes and explosion details
for a relatively low-mass core-collapse supernova.  Its age (as observed) and location out of the Galactic plane
suggest that the ejecta are not heavily contaminated by swept-up interstellar material.  Furthermore, numerous
measured electron temperatures in the range 7,000-18,000~K (e.g., Woltjer 1958, Miller 1978, Fesen \& Kirshner 1982, 
MacAlpine et al. 1989, 1996, and Satterfield 2010) imply that the line-emitting gas can be usefully analyzed
in terms of ionization, temperature, and chemical abundances using powerful photoionization codes.  Early
photoionization modeling endeavors (e.g., Henry \& MacAlpine 1982, Pequignot \& Dennefeld 1983) were reasonably
successful at matching observed line intensities and theoretical gas conditions for individual filaments.
However, there were notable problems.  For instance, observed $[$N~II$]$~$\lambda$$\lambda$6548,6583 from some
of the gas was weaker than expected (see Wheeler 1978).  Also, $[$C~I$]$~$\lambda$$\lambda$9823,9850 emission 
as measured at a number of locations (e.g., Henry et al. 1984) was stronger than computationally predicted
for reasonable carbon abundance. 

Regarding the $[$N~II$]$ problem, MacAlpine et al. (1996) employed long-slit spectroscopy, distributed
over the nebula, to show that there is a range of nitrogen emission from the gas, including regions with high implied 
nitrogen abundance as necessitated by CNO processing of H into He.  In addition, that study reported nebular locations
with low nitrogen and high sulfur abundances, suggestive of further processing including oxygen-burning.  
In a more comprehensive investigation,
MacAlpine et al. (2007) used optical and near-infrared long-slit spectroscopy to measure and correlate numerous 
emission lines at many locations in the nebula in order to obtain a better understanding of the overall range 
of nuclear processing, from the CNO cycle through helium-burning and nitrogen depletion, to extensive regions 
containing enriched products of oxygen-burning.  Then MS developed new photoionization calculations that do a 
better job of representing these various expected and observed stages of nuclear processing.  The improved
theoretical models also produce the observed $[$N~II$]$ and $[$C~I$]$ lines for reasonable nitrogen and carbon
abundances.

The next step in our on-going investigation of chemical abundances and nuclear processing in the Crab Nebula
is to map the overall distributions of elements like helium, carbon, nitrogen, oxygen, and sulfur.  
In this paper, we present new, flux-calibrated emission-line images obtained through custom-designed
interference filters and also element mass-fraction (abundance) maps that result from matching pixel-by-pixel
line fluxes with thousands of photoionization model calculations involving nested ranges of input abundances.
Much of this work was carried out as part of the undergraduate honors theses of Satterfield (2010) and Katz (2011).

\section{Observations}

During 2008 December~1-4, we imaged the Crab Nebula in He~II~$\lambda$4686, H$\beta$, He~I~$\lambda$5876, 
$[$O~I$]$~$\lambda$$\lambda$6300,6364, $[$N~II$]$~$\lambda$$\lambda$6548,6583 (including H$\alpha$, which was 
subsequently removed), $[$S~II$]$~$\lambda$$\lambda$6716,6731, and synchrotron continuum emission in roughly 
10-nm bandpasses centered at 5450 and 8050~\AA.  This was accomplished using the 2.7-m Harlan J. Smith Telescope 
at the McDonald Observatory, with the IGI imaging system and a 1024$\times$1024 TK4 CCD.  In addition, during 
2009 December~18-21, we used the 1.3-m McGraw-Hill Telescope at the MDM Observatory, with the 2048$\times$2048
Echelle CCD in direct mode, to image H$\beta$, $[$S~III$]$~$\lambda$9069, $[$C~I$]$~$\lambda$$\lambda$9823,9850,
and the aforementioned continuum bands in the nebula.  Additional direct-mode imaging was necessitated in the near-infrared
because of glass in the IGI camera at McDonald.  In all cases, the radiation was imaged through custom designed
interference filters manufactured by Omega Optical.  The filters, with transmission curves shown in Figure~1, 
were designed to transmit emission lines or their multiplets from gas expanding outwards at speeds up to
1700~km~s$^{-1}$.  In addition, at least two calibrated standard stars (selected from Feige~34, Feige~110,
G191-B2B, G158-100, and Hiltner~600) were observed through each filter.

\subsection{Initial Calibrations}

All images were bias-level subtracted using overscan regions and combined zero-second exposures, 
as accomplished with the {\it ccdproc} routine in the IRAF (Tody 1986) software package.
The optical images were flat-field corrected with the aid of dome flats, while the near-infrared images
were corrected for both flatness and fringing using sky flats.  Then all of the images were carefully registered
to coincide with the Crab Nebula Fabry-Perot datacube of Lawrence et al. (1995) (for reasons to be discussed later)
using the {\it geomap} and {\it geotran} IRAF routines.  Because of variable seeing and the fact that images 
would ultimately be directly compared with each other on a pixel-to-pixel basis, it was also necessary to
``degrade'' them to be consistent with the poorest seeing case, using {\it psfmatch} so stars would all 
have the same point-spread functions. Ultimately, we were able to median combine the multiple exposures
for each image without loss of data, with each pixel effectively representing the same gas in the different
images.  Resultant ``raw'' combined images are shown in Figures~2 and 3.

\subsection{Sensitivity and Flux Calibrations}

Next, we determined the system sensitivity for each filter and calibrated each image in terms of 
ergs~cm$^{-2}$~s$^{-1}$~pixel$^{-1}$.  Accurate and consistent sensitivity values were derived from the observed
standard stars at the McDonald Observatory, but the standard star results were not sufficiently consistent in the
MDM Observatory observations due to variable sky conditions.  Therefore, in the latter cases, we used the pulsar's
known energy distribution (Sollerman et al. 2000) for this purpose.

To obtain the system sensitivity, we followed a method similar to that described by Jacoby et al. (1987).  
Neglecting atmospheric absorption initially, the total flux (ergs~cm$^{-2}$~s$^{-1}$) from a standard star, S, incident
on the detector, through filter {\it i}, can be characterized by

\begin{displaymath}
F_S(i) = \int F_{S\lambda} T_{\lambda}(i) d\lambda
\end{displaymath}
which convolves F$_{S\lambda}$ (ergs~cm$^{-2}$~s$^{-1}$~\AA$^{-1}$) of each standard star above the atmosphere 
with each filter's transmission curve, T$_{\lambda}$({\it i}).  Then we can define a ``system sensitivity''
for each filter, taking into account atmospheric absorption, telescope optics, and detector response as

\begin{displaymath}
S(i) = F_S(i)/[C_S(i)\times10^{0.4kA(i)}]
\end{displaymath}
where C$_{S}$({\it i}) represents the observed star count rate and {\it k} is the appropriate airmass extinction 
coefficient (Massey \& Foltz 2000) in magnitudes per airmass A({\it i}).

Assuming uniformly distributed continuum emission across filter transmission bands, 
we can represent the measured continuum flux in any pixel with

\begin{displaymath}
F_C(i) = <S(i)>C_C(i)\times10^{0.4kA(i)}/<T_\lambda(i)>
\end{displaymath}
where $<S(i)>$ may be an average from several standard stars, $C_{C}$ is the continuum count rate for the bandpass 
and $<$T$_{\lambda}$({\it i})$>$ represents the effective average transmission for a filter.

Dealing with the line emission is more complicated.  For each pixel, and a given filter, there may be multiple
contributions from gas with different line-of-sight velocities and therefore different wavelength shifts
distributed over the filter transmission curve.  As seen in Figure~1, the transmission curves are not
perfectly uniform over the wavelength ranges where line emission may arrive.  Assuming various emission lines
have similar line-of-sight velocity distributions, we employed the Crab Nebula $[$O~III$]$-emission Fabry-Perot
datacube of Lawrence et al. (1995) to address this issue.  As discussed previously, all of our images have been
modified to the same resolution and registered to this 27-image datacube, which we used to produce an
appropriately-weighted wavelength distribution for line emission at each pixel.  Then, for the line flux
at each pixel, through a given filter {\it i},

\begin{displaymath}
F_L(i) = <S(i)>C_L(i)\times10^{0.4kA(i)}\sum[w_\lambda/T_\lambda(i)]
\end{displaymath}
where $\sum$$[$w$_{\lambda}$/T$_{\lambda}$({\it i})$]$ is the transmission weighted sum.  For any pixels
where $[$O~III$]$ emission in the Fabry-Perot datacube was not sufficient to yield reliable weighting factors,
the line flux was computed in a manner similar to that for the continuum, effectively centering the line
in the filter bandpass.

\subsection{Sky and Continuum Subtraction}

The flux-calibrated images (in ergs~cm$^{-2}$~s$^{-1}$~pixel$^{-1}$) still contained contributions
from both the background sky and synchrotron continuum.  To remove the former, we subtracted the average of 
numerous 300$\times$300 pixel areas in different directions and well outside the visible nebulosity.  Then, to ensure
consistency and mitigate signal-to-noise issues in the rest of the reduction process, we zeroed out any pixel
with a value lower than three times the remaining background's standard deviation above its average of zero.

For a given emission-line image, the closer in wavelength of the continuum images was used for removal
of the overlying synchrotron radiation.  To facilitate this process in an objective and consistent way,
we carefully identified the same regions in all images, devoid of line emission within the central part of the nebula;
and we subtracted the appropriate continuum image down to the previously-measured three standard deviations
above the background average of zero.

Flux-calibrated and continuum-subtracted images are shown in Figures~4 and 5.  Because the filter that isolated light
in $[$N~II$]$~$\lambda$$\lambda$6548,6583 also transmitted H$\alpha$~$\lambda$6563, our calibrated H$\beta$ image
was scaled by a factor of 2.85 and subtracted from the $[$N~II$]$ image in order to remove contributions
from H$\alpha$.  Also, as discussed by Davidson et al. (1982) and by Blair et al. (1992), He~II~$\lambda$4686
emission may arise in more diffuse gas compared with that producing other lines being considered here,
so images for that line in Figures~2 and 4 are presented as observational data only and will not be used 
for further comparative abundance analyses in this paper.

\subsection{Ratios of Emission Lines to H$\beta$}

To facilitate comparisons of our observed emission-line fluxes with output from the numerical photoionization code
Cloudy (Ferland et al. 1998), we could deal directly with the fluxes and then ratio results for different elements
in order to derive relative abundances, or we could normalize initially by dividing each line image by
the H$\beta$ image and then compare the observations with {\it relative} line intensities from the models.
We chose the latter approach.  Before creating the images for lines/H$\beta$, we reassigned an arbitrarily large
number (10$^{6}$) to each zero-valued H$\beta$ pixel as well as to each pixel with a flux value less than 10$^{-16}$
ergs~cm$^{-2}$~s$^{-1}$.  This eliminated the potential for infinite ratio values and for
boosting by more than a factor of ten resulting from division by H$\beta$.  The flux ratio images, corrected
for interstellar reddening according to E$_{B-V}$ = 0.47 and R = 3.1 (see Davidson \& Fesen 1985),
are presented in Figure~6.

\section{Element Mass Fraction Distributions}

Element mass-fraction maps were produced by comparing the observational flux ratio distributions of Figure~6
with a grid of 19,440 Cloudy code models involving nested element abundances (to be discussed below).
We developed algorithms that interact with the code output of computed emission-line ratios and identify
the model with the ``optimal-fit'' abundances for each image pixel.  Before elaborating further on this procedure
and showing results, we consider some caveats to keep in mind for computations of this type applied to the
Crab Nebula.

\subsection{Potential Complicating Factors}

The photoionization code assumes ionization and thermal equilibrium for the
line-emitting gas, whereas some particularly low-electron-density regions in the Crab Nebula
may not have had time to achieve equilibrium during the observed lifetime of the
remnant (see Davidson \& Tucker 1970).  On the other hand, the only emission
considered here that may be significantly affected in this regard is $[$O~I$]$~$\lambda\lambda$6300,6364,
and even these lines are produced in gas with recombination times roughly a sixth of
the observed age of the nebula.  All the other lines under consideration develop in gas
with high enough electron density for equilibrium to be reasonably assured (MS). 

In addition, as first emphasized by Davidson (1973), line-emitting gas in the Crab
Nebula may be immersed in synchrotron radiation approaching from all directions,
whereas the Cloudy code is used in a manner more appropriate for ionizing radiation
from one direction.  To consider this, MS developed algorithms for converting plane-parallel results to
what could be expected for convex cylindrical and spherical clouds in the synchrotron
nebula.  Changes of 30-40\% could result for computed $[$S~II$]$~$\lambda\lambda$6716,6731/H$\beta$ or
$[$O~I$]$~$\lambda\lambda$6300,6364/H$\beta$ in the most extreme cases of going from plane-parallel to
spherical geometries. Although we discuss below how this could affect our results,
we believe it is reasonable to consider plane-parallel geometries in the current
investigations of large-scale element distributions. 

Other potential complicating issues involve assumptions about model input
parameters such as ionizing radiation spectral characteristics, radiation flux striking
a cloud, and gas density.  As noted by Davidson \& Fesen (1985), Hester (2008), and
Charlebois et al. (2010), these conditions could vary spatially within the nebula.
Furthermore, our emission-line imaging technique involving relatively broad-bandpass
filters may lead to a variety of observed conditions along any line of sight.
On the other hand, MS were able to use similar inputs for parameters in developing
numerical models that gave reasonable matches to observational data for different
regions in the Crab Nebula.  For the present work, we tried to be as straightforward
as possible in developing a necessarily fairly crude two-dimensional overall
understanding of element distributions.  We always used the input ionizing
radiation spectrum adopted by MS, which is consistent with observations and
essentially the same as suggested in the documentation that accompanies the Cloudy
code.  Furthermore, we did not vary the flux of ionizing radiation striking gas clouds. 
With regard to density considerations, we examined a number of significantly
different potential paradigms, two of which we will discuss and compare here. 

\subsection{Matching Theoretical Models With Observed Pixels}

For our assumptions regarding the radiation field and density, we developed a
large grid of numerical photoionization simulations with extensive ranges of element
abundances.  The models were arranged in a nested hierarchy, beginning with six
helium abundances, then six carbon abundances for each helium abundance, six
nitrogens for each carbon, six oxygens for each nitrogen, and fifteen sulfurs for each
oxygen (2.5$\times$6$^{5}$ = 19,440 models). Specifically, letting MF(E)/MF(E)$_{\odot}$
be a modeled element (E) mass fraction compared with its solar value in the Cloudy code file,
we considered MF(He)/MF(He)$_{\odot}$ = 1.9, 2.8, 4.8, 9.8, 21, and 33.  For each helium value,
we let MF(C)/MF(C)$_{\odot}$ = 1.0, 6.0, 13, 19, 25, and 31.  Then for each carbon value,
MF(N)/MF(N)$_{\odot}$ = 0.1, 1.0, 1.9, 2.8, 3.7, and 4.6.  For each nitrogen value,
MF(O)/MF(O)$_{\odot}$ = 1.0, 7.0, 13, 20, 26, and 31.  Finally, for each oxygen value,
MF(S)/MF(S)$_{\odot}$ = 1.0, 2.0, 3.0, 4.0, 5.0, 6.0, 7.0, 8.0, 9.0, 10, 11, 12, 13, 14, and 15.
We used a finer abundance grid for sulfur for reasons to be discussed below.

In the Cloudy code, input abundances for each element (E) were specified as scale factors
\begin{displaymath}
{\it n} =[N(E)/N(H)]/[N(E)/N(H)]_{\odot},
\end{displaymath}
representing the number density compared with hydrogen divided by the solar number density compared with hydrogen.
To derive {\it n} for each MF(E)/MF(E)$_{\odot}$ above, it may be shown that
\begin{displaymath}
{\it n} = [MF(E)/MF(E)_{\odot}][MF(H)_{\odot}/MF(H)],
\end{displaymath}
where we let $[$MF(H)$_{\odot}$/MF(H)$]$ = 0.7/$[$1-MF(He)-MF(heavier elements)$]$.

For a given element, the total mass-fraction range considered was adopted
by comparing models with observations, and there is correspondence with
theoretical mass fractions to be expected in the core of an evolved 10~M$_{\odot}$ star (see
Woosley \& Heger 2007).  We did not actually consider the highest possible mass fractions suggested
by the stellar models because computed line-emission ratios became less sensitive to changes above
our upper abundance cutoffs.  Other related element abundances were also varied as
appropriate; for instance, silicon and argon mass-fractions followed those of sulfur,  
since they are all produced from oxygen-burning. Computed line intensities were
extracted from the models at a distance into a cloud of three times the so-called
Str\"{o}mgren length at which half the atomic hydrogen is ionized; and it was found that
gas deeper into a cloud does not contribute substantially more line emission. 

In order to find optimal matches to imaged pixels from among the 19,440 abundance models,
we considered several different approaches, ultimately adopting one that matches helium first, rather 
than giving equal weight to all elements simultaneously. This took into account helium's
importance for physical conditions like the ionization structure.
For each pixel (representing the same nebular gas for all images; see
\S~2.1), a numerical algorithm was used to select the subset of 19,440/6
models with a helium abundance (from the six considered) most closely matching the observed He~I~$\lambda$5876/H$\beta$
ratio.  From those models, the smaller subset of 19,440/36 models with carbon abundance most
closely matching observed $[$C~I$]$~$\lambda\lambda$9823,9850/H$\beta$ was selected, followed by similar
procedures for nitrogen, oxygen, and sulfur (involving only 15 models in the last case).  
Ultimately, one computed model was identified as having abundances most closely representing each image pixel.

The computed and observed line ratios, as matched for initially establishing abundances at each pixel,
had average residuals of 10-38\% compared with the measurements.  However, the ultimately selected abundance model
for each pixel will not necessarily have the same lines that were matched to set up the abundances.
For instance, once a helium abundance is initially selected, there are 19,440/6 or 3,240 models for that abundance,
any of which could ultimately represent the pixel, depending on line matches and abundances for the other elements.
On the other hand, it was reassuring that the helium line residuals for the {\it final} models of all the pixels
averaged only 23\% compared with the observations.

While investigating details of the matching process, it was noted that, for some
circumstances, the computed $[$S~II$]$~$\lambda\lambda$6716,6731/H$\beta$ ratio first increased with rising
sulfur mass fractions up to about 6-8 times solar, and then flattened or even decreased
somewhat.  The latter situation could lead to redundancy for some pixels, whereby
their matched observed and theoretical line ratios might correspond with either high
or low sulfur abundance. Experimentation showed that this potential double-value
dilemma is more likely to occur with lower helium abundances (due to effects on
ionization structure) and for lower hydrogen densities.  Plotting of temperature,
density, and line emissivity suggests it is a temperature effect, whereby increasing the
modeled sulfur and silicon abundances results in more collisional cooling, partly from
the fine-structure $[$S~III$]$ and $[$Si~II$]$ transitions discussed by Henry (1993).  As the
sulfur mass fraction increases and the gas temperature goes down, the $[$S~II$]$/H$\beta$ ratio
first increases because of more S$^{+}$ and then declines because of relative temperature
dependences of $[$S~II$]$ and H$\beta$ emission.  To deal with this dilemma, we used a finer
abundance grid for sulfur (fifteen abundances rather than six as for other elements); and our algorithm that
matches models with observations was set to flag any potential double-valued pixels
and to select the lower abundance if two are possible for an observation.  The problem
is significantly less important for $[$S~III$]$~$\lambda$9069 emission.
Whereas the $[$S~II$]$ and $[$S~III$]$ lines lead to similar sulfur abundance spatial distributions,
$[$S~III$]$ emission yields regions with higher sulfur content (places where
the $[$S~II$]$ algorithm detected two abundance possibilities and selected the lower one). 
Because of this situation, {\it sulfur abundance maps derived from $[$S~III$]$ emission should
be considered as more realistic}. Although other element abundances were
derived from models with a range of sulfur and silicon content, their mass-fraction
maps were not significantly affected by this problem.

It could be useful to test the validity of the pixel-model matching process by comparing our theoretical
computations with published spectroscopic measurements for {\it multiple} ionization stages of oxygen 
at various positions.  Unfortunately, this becomes seriously complicated when trying to match spectroscopic 
slits or apertures with our image pixels and also when considering imaged emitting gas with different 
velocities along the line of sight.  
Our best opportunity for this type of test may be position 8 of Fesen \& Kirshner (1982), which 
appears to be relatively well defined with a dominant single-velocity spectrum.  Our models averaged 
over four pixels at that location give $[$O~II$]$~$\lambda$3727/H$\beta$ = 11.83 (compared with 14.90 measured
by Fesen \& Kirshner) and $[$O~III$]$~$\lambda$5007/H$\beta$ = 7.66 (compared with 9.82 measured by Fesen \& Kirshner).
The residuals are consistent with the averages found above from matching our model outputs with our imaged lines.

Carbon also provides observed lines representative of multiple ionization stages,
namely $[$C~I$]$~$\lambda\lambda$9823,9850 as imaged and modeled here and also C~III$]$~$\lambda$1909
and C~IV~$\lambda$1549 in the ultraviolet.  However, as discussed by Davidson (1978), 
Davidson et al. (1982), Henry \& MacAlpine (1982), and Blair et al. (1992), much of this
ultraviolet line emission probably comes from lower-density, more diffuse gas, compared with the bright 
optical-line-emitting condensations being considered here.   

As noted previously, for this investigation we will examine two sets of
photoionization model simulations, with different assumptions regarding gas density
input parameters.  The first models are similar to those of MS, with constant
hydrogen density of 3,000~cm$^{-3}$ and constant log U = -3.5 (where the so-called
ionization parameter, U, is defined for the Cloudy code as $\Phi/cN$, or the flux of
hydrogen ionizing photons striking a cloud divided by the speed of light times the
hydrogen density). With these assumptions, the total nuclear density can be higher
by as much as a factor of 4-5 for regions with very high helium abundances, but this is
well within the density variation range suggested by Jun (1998).  The second grid of models is
characterized by {\it constant nuclear density} of 6,000~cm$^{-3}$, resulting in a range of 4-5 for
hydrogen density as the helium abundance varies.  Since both types of simulations
assume the same ionizing radiation flux, the second set also involves appropriately
varying ionization parameters.  The specific density values in each case were considered because they produce similar 
electron-density-dependent $[$O~II$]$~$\lambda$3726/$\lambda$3729 ratios, 
which are consistent with observations (e.g., Fesen \& Kirshner 1982).

\subsection{Mass-Fraction Plots}

The resultant element abundance plots for the two density paradigm cases described above are
shown in Figures~7 and 8.  Helium is presented in terms of its overall mass percentage in
the emitting gas, whereas other elements are given as rough multiples of their solar mass fractions.  
 
We note that the helium mass fraction maps are virtually indistinguishable for the
two input density regimes.  Helium comprises about 40\% to $>$90\% of the emitting
gas in different regions, with an overall derived distribution very much like that reported
by Uomoto \& MacAlpine (1987).  There is an area with relatively low helium in the
north, possibly indicative of an ambient interstellar cloud (Morrison \& Roberts 1985, Fesen \& Gull 1986).
In addition, the roughly east-west high-helium band or torus (see also Lawrence at al.
1995) stands out as a prominent feature. 

For the two types of models illustrated, the carbon, nitrogen, and oxygen mass-fraction 
maps are similar in terms of overall structure, with some quantitative
differences especially for carbon. On the other hand, the sulfur maps show some notable
variation.  As discussed previously, double valuing tends to occur for
matching $[$S~II$]$/H$\beta$ emission and sulfur abundance, resulting in the lower possible
abundance being selected by our algorithm.  Therefore, the sulfur maps derived from
$[$S~II$]$ emission show some unrealistically low mass-fractions.  We do not know which (if either)
set of input density assumptions is more appropriate for Crab Nebula emitting gas, and we will {\it arbitrarily}
concentrate on the constant nuclear density maps (Fig.~8) in further discussion below. 

Next, we consider carbon, which is a product of helium fusion.  The distribution
map of Figure~8(b) shows a wide range of carbon mass fraction, from roughly solar to
more than 30 times solar.  This is consistent with the results of MS, who demonstrated how the
strong observed $[$C~I$]$~$\lambda\lambda$9823,9850 doublet arises from electron collisional excitation
in gas with very high helium abundance and who then used spectroscopic measures along with photoionization simulations
to derive carbon concentrations more than 10 times solar in parts of the Crab Nebula.
As shown by the ionization, temperature, and density diagrams in Figures 2-5 of MS, very high helium abundance 
causes rapid depletion, near the face of a cloud, of ionizing photons with energy above 24.6~eV, 
the ionization potential of He$^{0}$.
Because C$^{+}$ has a comparable ionization potential and that for C$^{0}$ is below 13.6~eV,
the dominant ionization stage for carbon is C$^{+}$.  Recombination of C$^{+}$ to C$^{0}$ in
the warm He$^{0}$ and H$^{+}$ zone leads to strong collisionally-excited $[$C~I$]$~$\lambda\lambda$9823,9850 emission,
which becomes more pronounced for higher helium abundance as illustrated in Figure~9 (from Katz 2011).

We note that the carbon mass-fraction distribution does not exactly follow the  
observed $[$C~I$]$ flux of Figure~5(h) or the $[$C~I$]$/H$\beta$ map of Figure~6(f).  
This can be understood from examination of the helium distribution in Figure~8(a) and
consideration of Figure~9. Regions with relatively lower $[$C~I$]$ emission and lower
helium abundance may actually contain more carbon compared with regions of
higher $[$C~I$]$ emission and higher helium abundance.  

Helium abundance is not the only reason why $[$C~I$]$~$\lambda\lambda$9823,9850 emission 
is stronger in the Crab Nebula than in other types of astronomical objects.  
Because of effects on photon mean free paths and ionization transition zone widths, the
ionization parameter and gas density can also play roles in producing this emission, as
illustrated in Figure~10 (from Katz 2011).  Planetary nebulae or H~II regions may be characterized by
gas with log~U of order 0 (e.g., Becerra-Davila et al. 2001), so we would not
expect strong $[$C~I$]$/H$\alpha$ from them, even if the helium abundance were high.  In
addition, for expected densities of 10$^{6}$~cm$^{-3}$ or higher, $[$C~I$]$ emission should not
stand out in a Seyfert galaxy or quasar spectrum.

As shown in Figures~5(e) and 6(c), $[$N~II$]$$\lambda\lambda$6548,6583 emission is strong in the
northern part of the nebula, in accord with the long-slit spectroscopic findings of
MacAlpine et al. (1996).  Resultant derived elevated nitrogen abundances could have arisen 
from the CNO-cycle, with relatively low helium in that region indicative of interaction with an ambient cloud (see above).
MS showed how high helium abundance is conducive to producing strong $[$N~II$]$ emission,
so lower helium abundance in the north (see Figure~8(a)) also leads to higher derived
nitrogen abundance for a measured $[$N~II$]$ intensity.  According to Figure~8(c),
the nitrogen abundance tends to be lower in those parts of the nebula where a lot of carbon
exists, probably as a result of
$^{14}$N($\alpha$,$\gamma$)$^{18}$F($\beta$$^{-}$$\nu$)$^{18}$O($\alpha$,$\gamma$)$^{22}$Ne     
processing which would  
be expected to accompany helium-burning (Nomoto 1985).  Infrared neon lines have been measured in the Crab Nebula
by Temim et al. (2006). They reported strong emission from $[$Ne~II$]$~12.8~$\mu$m and $[$Ne~III$]$~15.5~$\mu$m
near the center of the nebula, where we find relatively low nitrogen abundance.  Furthermore, MacAlpine et al. (2007)
found an inverse relation between their measured $[$N~II$]$~$\lambda$6583 and the Temim et al. 
measured $[$Ne~II$]$~12.8~$\mu$m for a particular position.  Finally, we note that the apparent high
nitrogen abundance around some edges of the nebula may be real, or it could be due
in part to our use of $[$N~II$]$/H$\beta$ ratios for regions with very weak H$\beta$ emission
(compare Figures~4(b), 5(e) and 6(c)).

Turning our attention now to the observed $[$O~I$]$~$\lambda\lambda$6300,6364 fluxes, $[$O~I$]$/H$\beta$
ratios, and oxygen mass-fraction distributions of Figures~4(d), 6(b), and 8(d),
respectively, we see strong emission and high oxygen abundance in the central 
east-west band of the nebula, including in outlying parts of the high-helium torus
region.  Oxygen could arise from helium or carbon-burning. From examination of
element overlaps (or lack thereof) in the Woosley \& Heger (2007) 10~M$_{\odot}$ stellar core mass-fraction
diagram, it may seem odd that some of the highest oxygen abundances in Figure~8(d)
appear to coincide with regions where the helium abundance is also quite high. 
However, this apparent anomaly could result from gas mixing in the Crab Nebula or
from a combination of different gas regimes along our line of sight. Mixing and
averaging of emission along sight lines may also help to explain why we do not
derive oxygen mass fractions as high as the maximum in the Woosley \& Heger
mass cut simulation.  In addition, as noted in \S~3.1 above, for $[$O~I$]$ emission,
consideration of spherical geometry cloudlets immersed in the synchrotron radiation
field could lead to a decrease of as much as 40\% in the computed $[$O~I$]$/H$\beta$ ratio and
to a corresponding increase in the derived oxygen mass fraction for a given pixel.
Going to a cylindrical filamentary geometry would have somewhat less effect.

Finally, we consider the $[$S~II$]$~$\lambda\lambda$6716,6731 and $[$S~III$]$~$\lambda$9069 
emission in Figures~5(f), 5(g), 6(d), and 6(e), along with the $[$S~III$]$-derived (see previous discussion) sulfur
mass-fraction map of Figure~8(f).  The general $[$S~II$]$ emission features are similar to
those in MacAlpine et al. (1989) and Charlebois et al. (2009).  The raw $[$S~III$]$ image
here has less signal-to-noise, is more dominated by continuum, and shows fewer
prominent features compared with the raw $[$S~II$]$ image.  However, the flux-calibrated   
and continuum-subtracted images look very much alike.  We note that $[$S~III$]$~$\lambda$9069
was chosen for this investigation, rather than the stronger line at $\lambda$9531 because of
telluric water vapor absorption in the region of the latter (see Stevenson 1994, Vermeij et al. 2002, 
MacAlpine et al. 2007). As illustrated in Figure~8(f), high sulfur abundance regions tend to avoid
some western parts of the nebula, being concentrated primarily to the east, as might
be understood in terms of off-center oxygen ignition (see Woosley \& Weaver 1986). 
The sulfur concentrations appear to avoid regions with the highest helium and
oxygen abundances, instead occupying the apparent loop-like structures.  
Fairly high sulfur in the south-east edge filaments
could be real, or it may result from weak H$\beta$ in the $[$S~III$]$/H$\beta$ ratios used for
comparison with the model computations.  

\section{Summary} 

We have presented new raw and flux-calibrated, continuum-subtracted
emission-line images of the Crab Nebula in He~II~$\lambda$4686, H$\beta$, He~I~$\lambda$5876,
$[$O~I$]$~$\lambda\lambda$6300,6364, $[$N~II$]$~$\lambda\lambda$6548,6583, $[$S~II$]$~$\lambda\lambda$6716,6731,
$[$S~III$]$~$\lambda$9069, and $[$C~I$]$~$\lambda\lambda$9823,9850.  
Ratios of line images to H$\beta$ were compared with an extensive grid of
photoionization simulations in order to derive element mass-fraction maps
for helium, carbon, nitrogen, oxygen, and sulfur.  These distributions should be
considered as somewhat simplified approximations in the sense that plane-parallel
geometry was assumed, along with global assumptions regarding ionizing
radiation and gas physical conditions.  It may be possible in future studies to divide
the nebula into different regions with separate assumptions, but we
believe from considering various parametric schemes, that the results presented are
representative of the Crab Nebula's general characteristics.  Although there has
apparently been mixing physically as well as observationally along the line of sight,  
and also pulsar wind-driven material, the element distributions are distinct and
informative. They illustrate regions of gas in which various stages of nucleosynthesis
have occurred, including the CNO-cycle, helium-burning, carbon-burning, and
oxygen-burning.  It is hoped that the calibrated observations and chemical abundance
distribution maps will be useful for developing a better understanding of the precursor
star evolution and the supernova explosive process. Work currently underway will
extend this investigation to derive actual {\it mass} distributions and estimates of overall
masses for the various elements. 

\acknowledgments

This work was supported by Trinity University and the endowed Charles A. Zilker Chair position.
We also thank the staffs of the McDonald Observatory and the MDM Observatory for providing technical assistance.

\clearpage

\begin{figure}
\epsscale{1.0}
\plotone{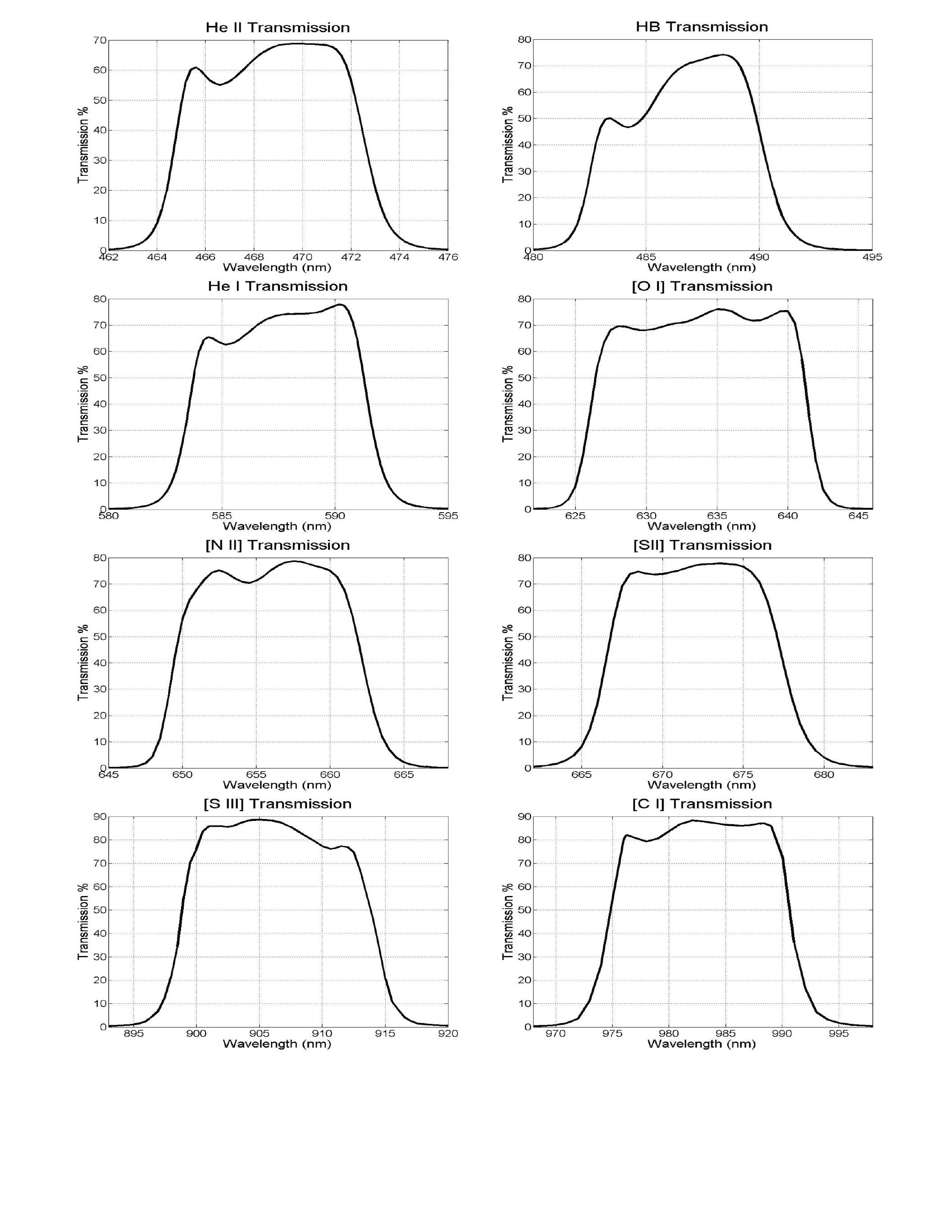}
\caption{Filter transmission curves.
 \label{fig1}}
\end{figure}
\clearpage

\begin{figure}
\epsscale{1.0}
\plotone{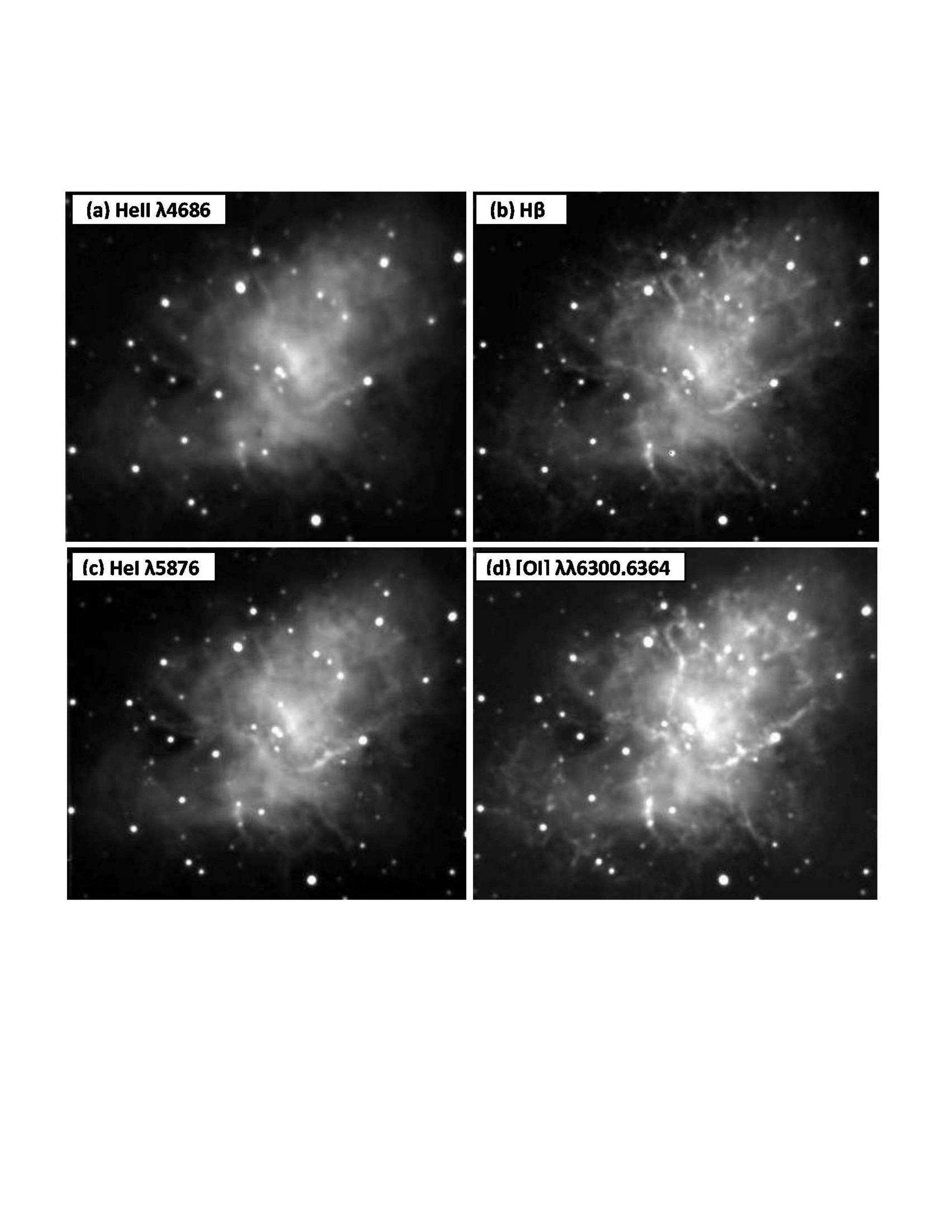}
\caption{Raw emission-line images.
 \label{fig2}}
\end{figure}
\clearpage

\begin{figure}
\epsscale{1.0}
\plotone{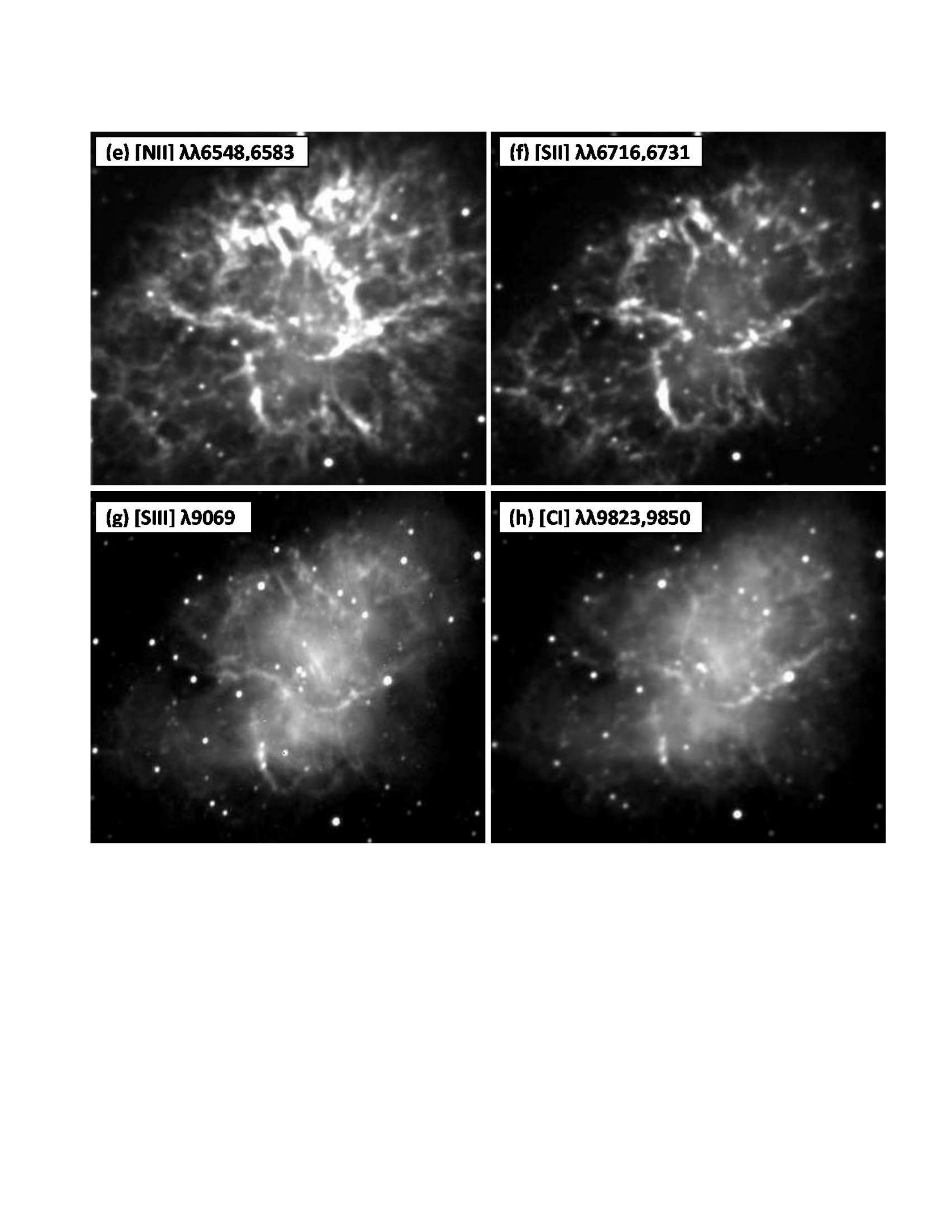}
\caption{Raw emission-line images continued.
 \label{fig3}}
\end{figure}
\clearpage

\begin{figure}
\epsscale{1.0}
\plotone{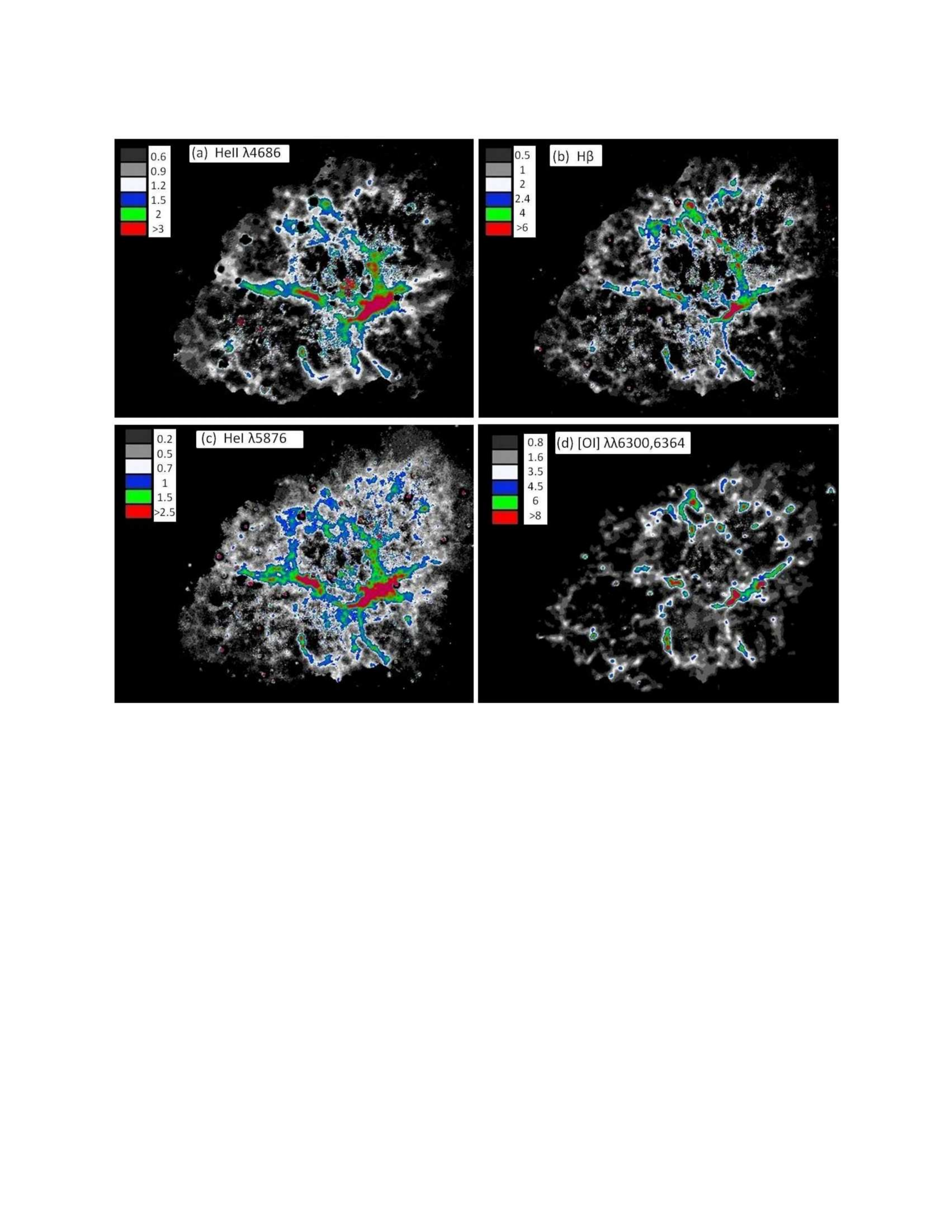}
\caption{Flux-calibrated, continuum-subtracted emission-line images.  Color coded fluxes are the numbers shown times 
10$^{-15}$~ergs~cm$^{-2}$~s$^{-1}$.
 \label{fig4}}
\end{figure}
\clearpage

\begin{figure}
\epsscale{1.0}
\plotone{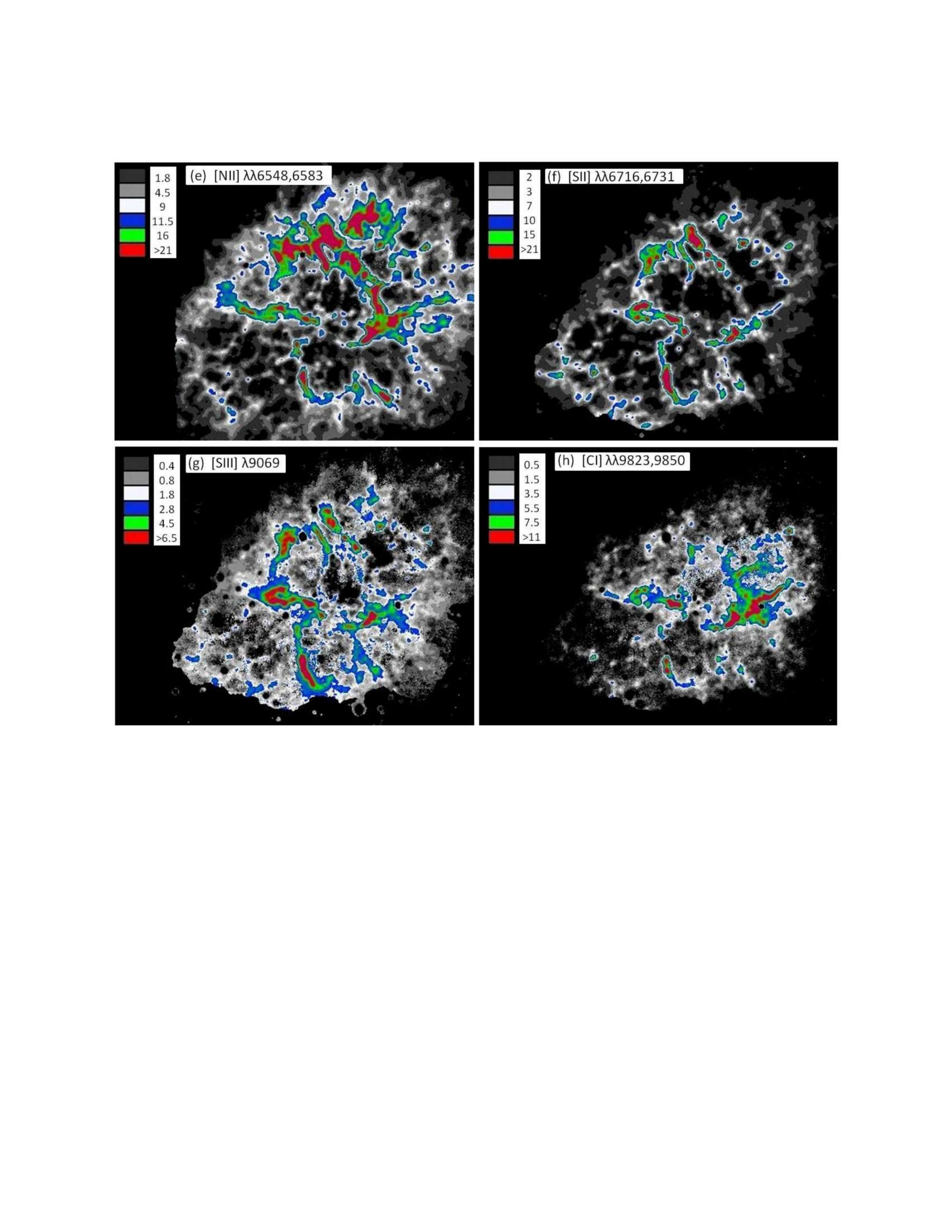}
\caption{Flux-calibrated, continuum-subtracted emission-line images continued.  Color coded fluxes are the numbers shown 
times 10$^{-15}$~ergs~cm$^{-2}$~s$^{-1}$.
 \label{fig5}}
\end{figure}
\clearpage

\begin{figure}
\epsscale{1.0}
\plotone{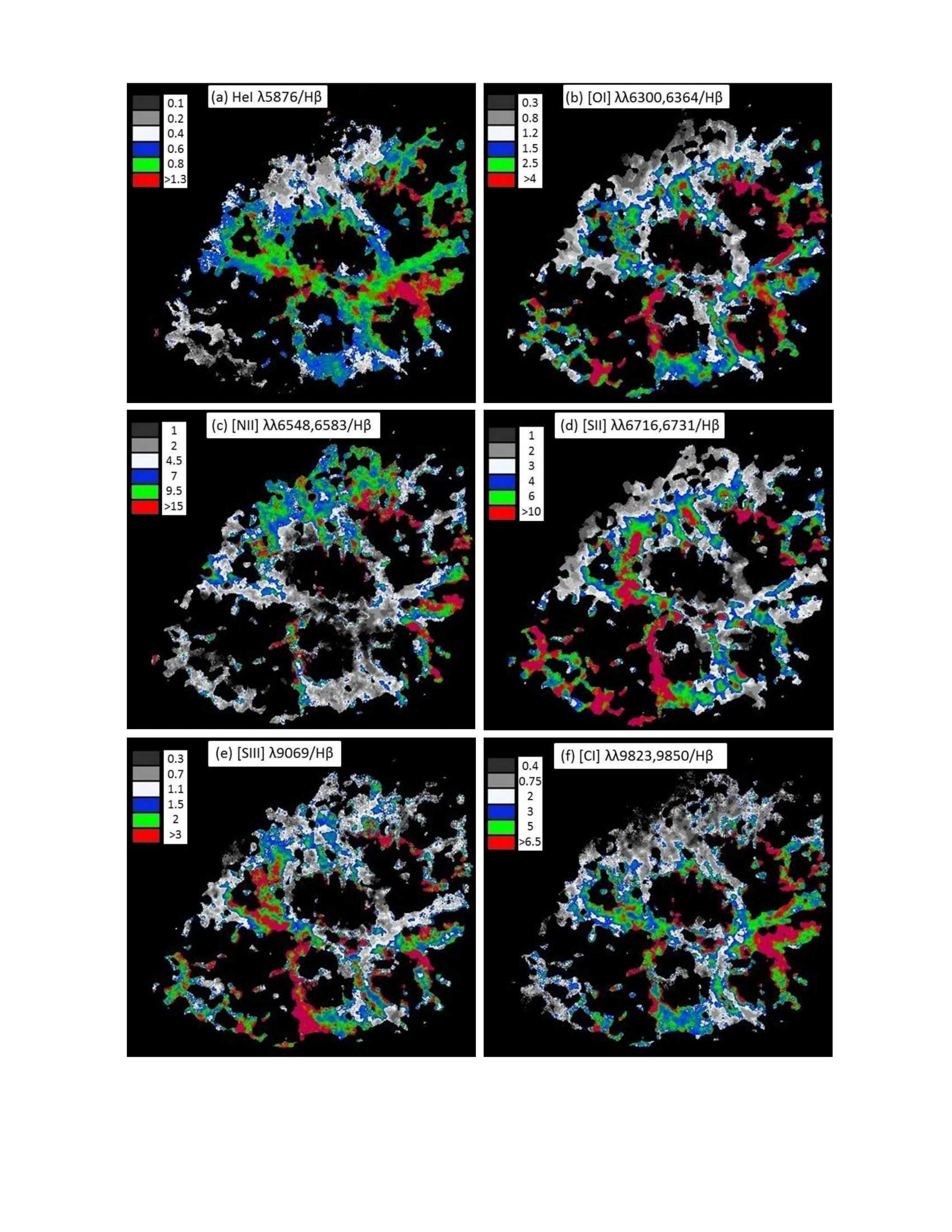}
\caption{Ratios of emission-line images to the H$\beta$ image.
 \label{fig6}}
\end{figure}
\clearpage

\begin{figure}
\epsscale{1.0}
\plotone{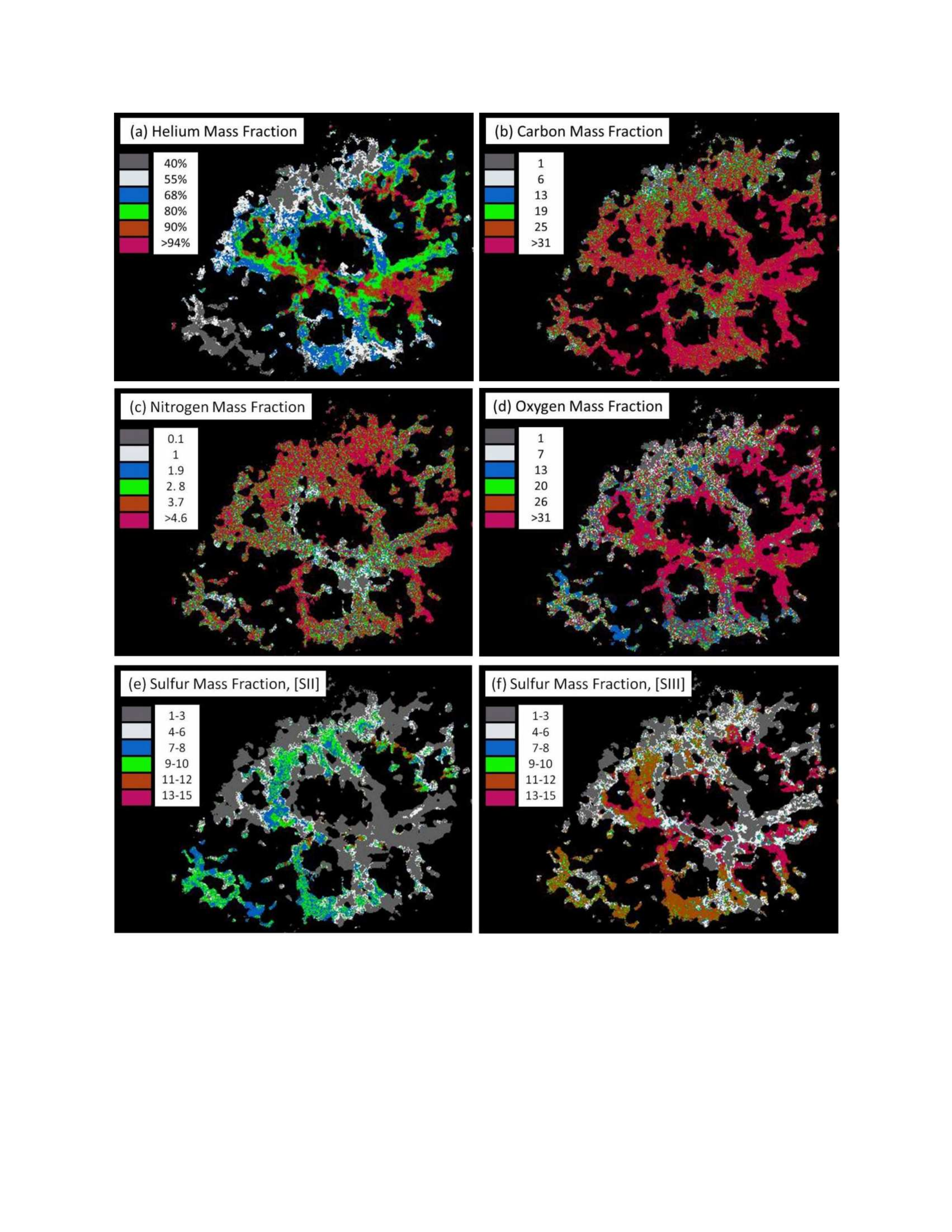}
\caption{Computed mass-fraction distributions for constant hydrogen density and constant ionization parameter 
(see text). Helium mass fractions are specified as overall percentage of emitting-gas mass, whereas other elements are 
labeled as multiples of their solar mass fractions. 
 \label{fig7}}
\end{figure}
\clearpage

\begin{figure}
\epsscale{1.0}
\plotone{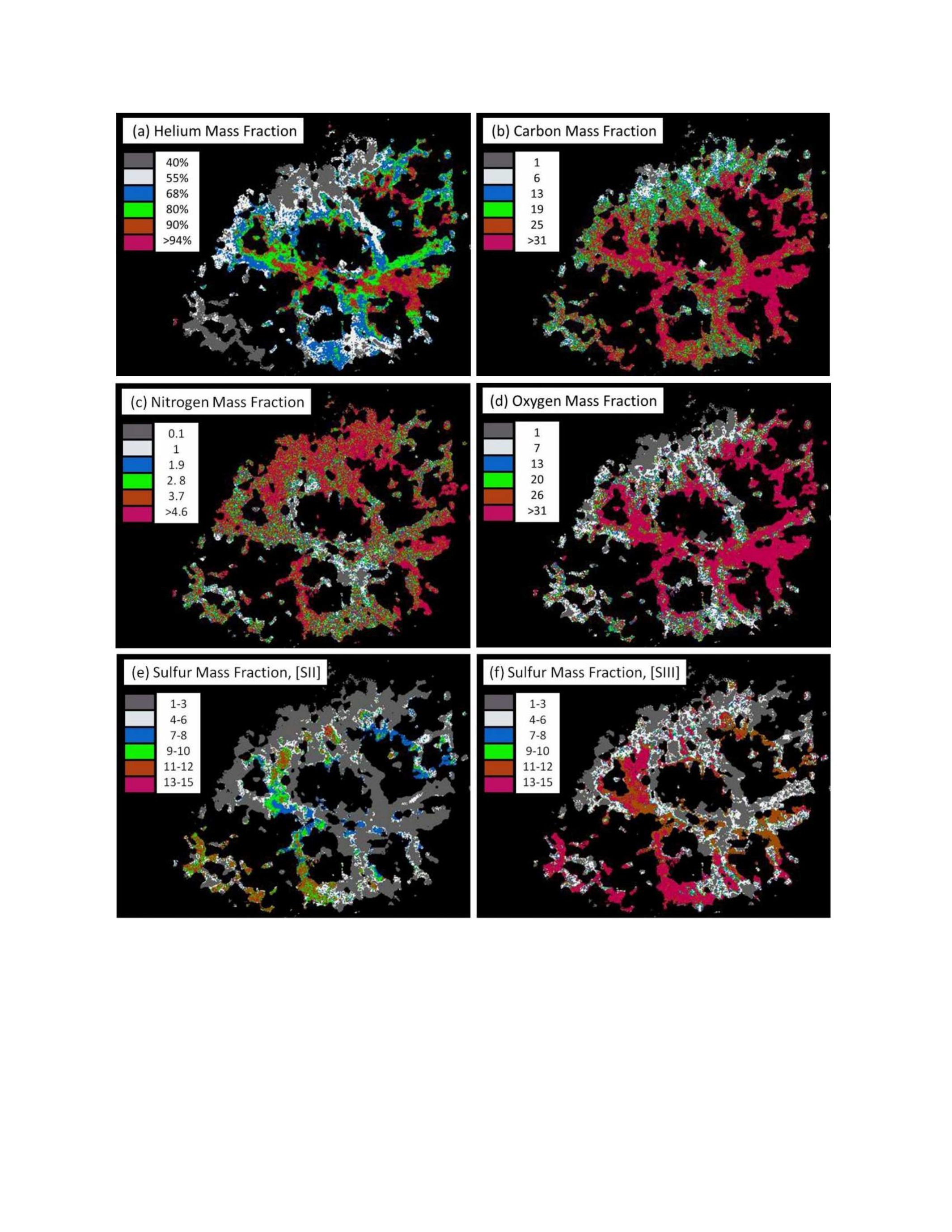}
\caption{Computed mass-fraction distributions for constant overall nuclear density, with varying hydrogen
density and ionization parameter (see text). 
Helium mass fractions are specified as overall percentage of emitting-gas mass, whereas other elements are labeled
as multiples of their solar mass fractions.
 \label{fig8}}
\end{figure}
\clearpage

\begin{figure}
\epsscale{1.0}
\plotone{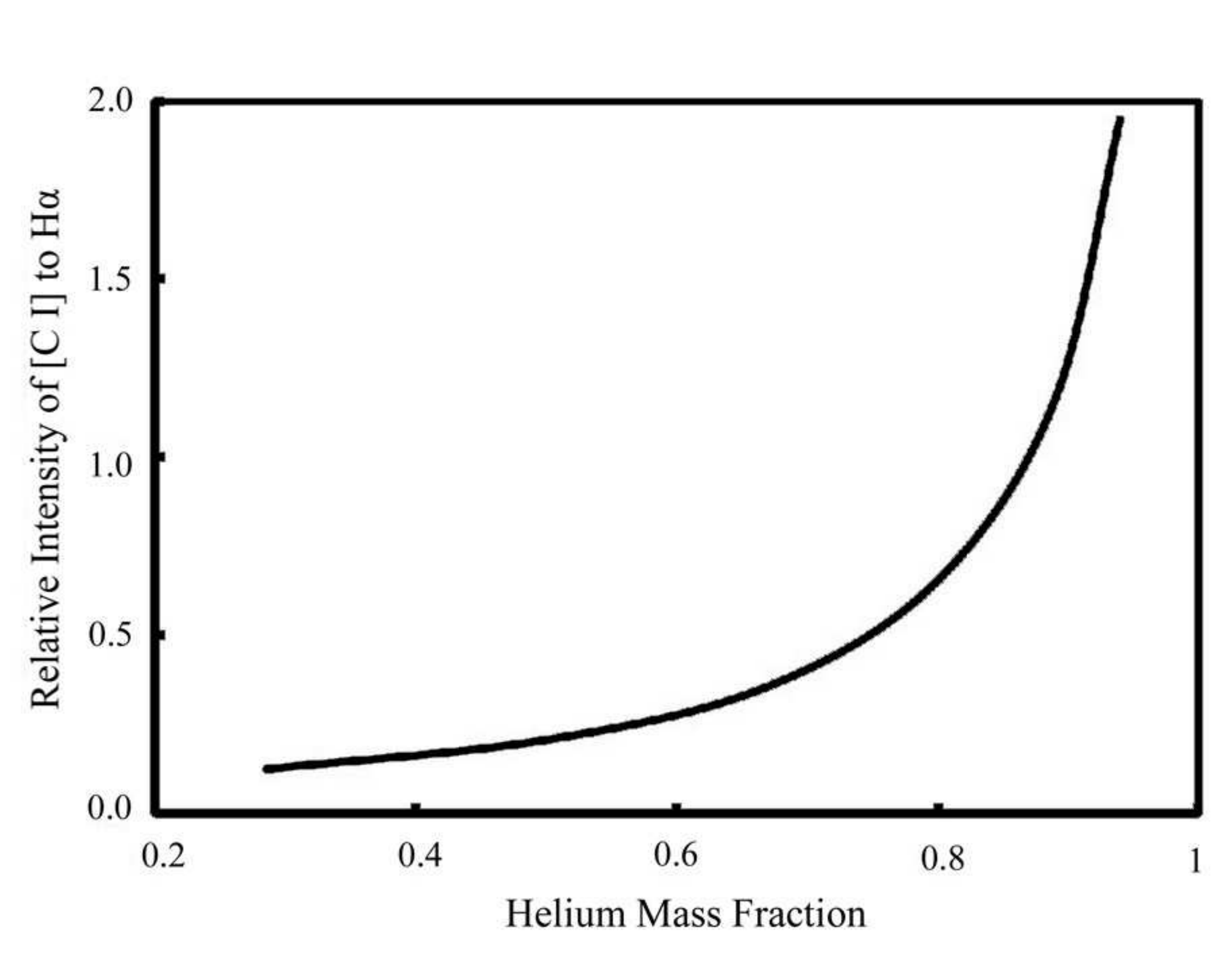}
\caption{Computed $[$C~I$]$~$\lambda\lambda$9823,9850 to H$\alpha$ ratio as a function of helium mass fraction.
For hydrogen density=3,000~cm$^{-3}$ and log~U=-3.5.
 \label{fig9}}
\end{figure}
\clearpage

\begin{figure}
\epsscale{1.0}
\plotone{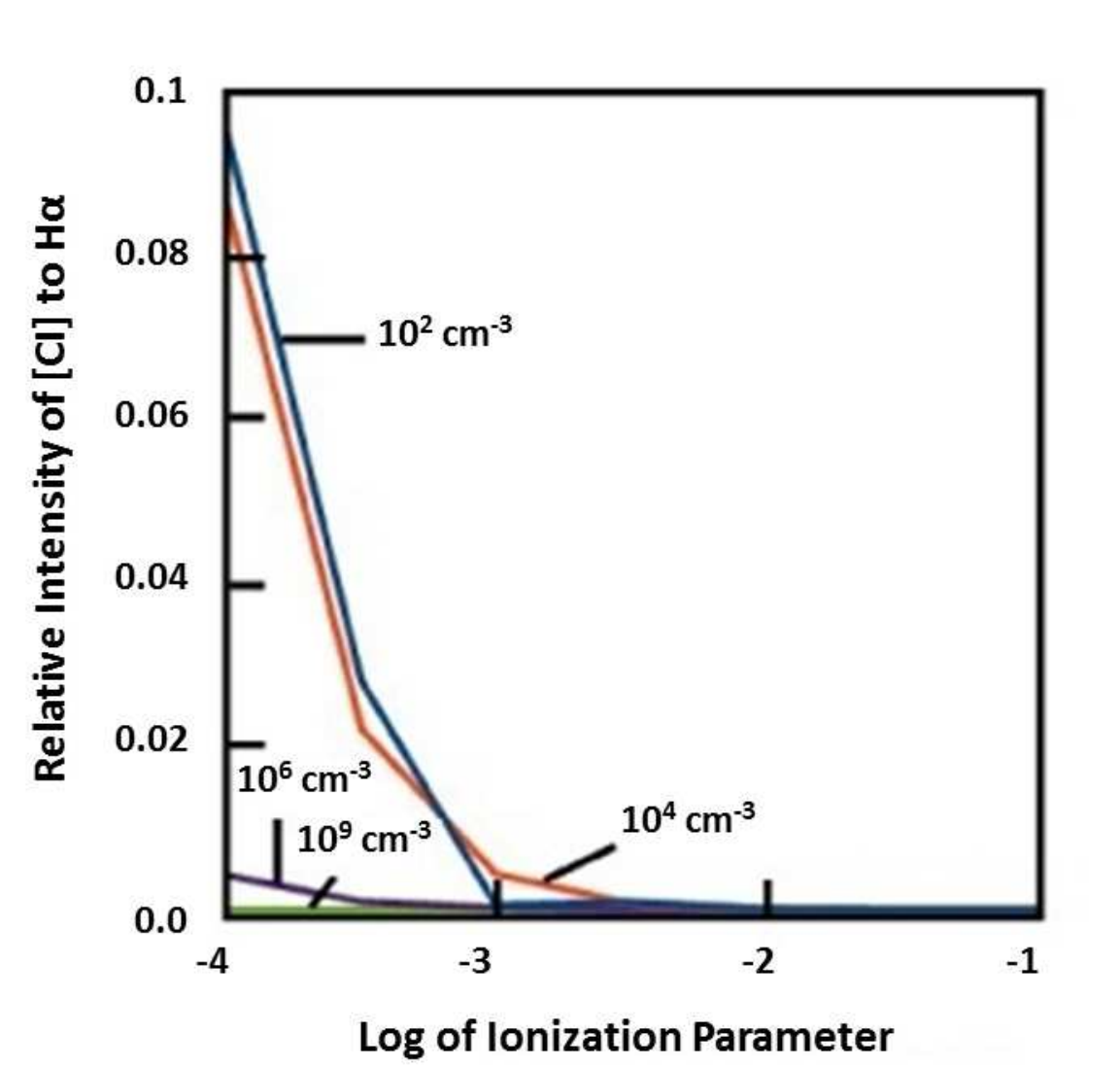}
\caption{Computed $[$C~I$]$~$\lambda\lambda$9823,9850 to H$\alpha$ ratio as functions of log ionization parameter 
and line-emitting gas density (cm$^{-3}$).  For solar element abundances.
 \label{fig10}}
\end{figure}
\clearpage

\end{document}